\documentclass[10pt,a4paper,reqno]{amsart}
\usepackage{amsthm,amssymb,amsmath,graphicx}
\usepackage{mathrsfs,setspace,pstricks,multicol,latexsym}
\usepackage{epsfig, subcaption, graphics}
\usepackage{hyperref}
\usepackage{multirow}
\usepackage{dsfont}
\usepackage{float}
\hypersetup{
	colorlinks,
	citecolor=blue,
	filecolor=black,
	linkcolor=red,
	urlcolor=black
}
\advance\oddsidemargin by -1.6cm
\advance\evensidemargin by -1.6cm

\numberwithin{equation}{section}

\newcommand\new[1]{}
\newtheorem{Theorem}{Theorem}[section]
\newtheorem{Definition}[Theorem]{Definition}

\newtheorem{Remark}[Theorem]{Remark}

\newcommand{\no}{\nonumber}

\newcommand{\ds}{\displaystyle}

\newcommand{\la}{\lambda}

\newcommand{\sig}{\sigma}

\let\emptyset\varnothing

\allowdisplaybreaks

\date{}

\begin{document}

\pagenumbering{arabic}
%\author[M. K. Das, A. Goswami]{Milan Kumar Das \affil{1}, Anindya Goswam \affil{2}\comma\corrauth}
%
%\address{\affilnum{1}\ Department of Mathematics, IISER Pune, India\\
%\affilnum{2}\ Department of Mathematics, IISER Pune, India}
%
%\emails{{\tt das.milan2@gmail.com}\ (M. K. Das), {\tt anindya@iiserpune.ac.in}\ (A. Goswami)}

\author{Milan Kumar Das}
\address{}
\email{das.milan2@gmail.com}
\thanks{The first author acknowledges SRF grant of UGC.}

\author{Anindya Goswami}
\address{IISER Pune, India}
\email{anindya@iiserpune.ac.in}

\title[Testing of Binary Regime Switching Models using Squeeze Durations]{Testing of Binary Regime Switching Models using Squeeze Duration Analysis}\thanks{This research was supported in part by the SERB MATRICS (MTR/2017/000543), DST FIST (SR/FST/MSI-105).}

%\addtocounter{footnote}{-1} \vskip 1 true cm
\begin{abstract}
We have developed a statistical technique to test the model assumption of binary regime switching extension of the geometric Brownian motion (GBM) model by proposing a new discriminating statistics. Given a time series data, we have identified an admissible class of the regime switching candidate models for the statistical inference. By performing several systematic experiments, we have successfully shown that the sampling distribution of the test statistics differs drastically, if the model assumption changes from GBM to Markov modulated GBM, or to semi-Markov modulated GBM. Furthermore, we have implemented this statistics for testing the regime switching hypothesis with Indian sectoral indices.
\end{abstract}

\maketitle

\section{Introduction}

In the work of Black, Scholes, and Merton in 1973, a geometric Brownian motion (GBM) model was used to explain the dynamics of risky asset prices. This was subsequently reexamined and extended by several authors to fill the gaps between the empirical observations and the consequences from theoretical models. In this context, stochastic models following the work of Heston (1993) \cite{heston} appear promising. However the calibration of such models and their extensions became increasingly challenging. Simultaneously the regime switching models following the works of Hamilton (1989) \cite{hamilton} and Di Masi et.al (1994) \cite{Di}, also earned popularity in the literature and research in this direction continue to grow. Regime switching GBM is one such extension. Various authors assumed regime switching models and addressed theoretical option pricing and portfolio optimization problems following \cite{Di}. For details, the readers may refer to  \cite{BAS}, \cite{BO}, \cite{FAN}, \cite{JA}, \cite{LI}, \cite{LIA}, \cite{PA} \cite{SIU}, \cite{SUX}, \cite{SW}, \cite{WEI}, \cite{YUE}. This list is merely indicative and not exhaustive by any means. However, the statistical verification of effectiveness of the regime switching models for an asset price data has not yet been satisfactorily addressed in the literature. Nevertheless, using a particular regime switching model, some stylized facts in the daily return had been explained in \cite{BU}, \cite{RY}. It is important to note that, in a regime switching dynamics, each of the key market parameters evolves as a finite state pure jump process which might or might not be a Markov chain. Replacement of a Markov chain by a semi-Markov chain results in an immediate extension. As per our knowledge, there is no comprehensive statistical analysis which helps discriminating among the cases of GBM, Markov-modulated GBM(MMGBM) and semi-Markov modulated GBM(SMGBM) for modeling a given asset price time series data. However, in \cite{BU} the authors have studied the stylized facts of returns of daily time series and have shown that the hidden semi-Markov models describe the stylized facts including the slow decay in the autocorrelation function (ACF) of the squared returns in a better way than what MMGBM model does.

The main difference between a Markov chain and a semi-Markov chain is in its instantaneous transition rate. It is just a constant matrix for Markov case whereas for the semi-Markov chain it is a matrix valued function on the positive real numbers. Despite the presence of many computational and theoretical studies on the SMGBM model for asset prices, the use of SMGBM is not as popular as it deserves to be. In recent works \cite{AGMKG}, \cite{AJP}, \cite{AS}, \cite{MAT} it is shown that the option price equation in SMGBM model can be solved with similar ease to the MMGBM counterpart. Furthermore, the portfolio optimization problem in SMGBM leads to the same optimal strategy as in the MMGBM case \cite{MAN}. Hence the SMGBM model is as appealing as the special case of MMGBM in terms of the mathematical tractability. So far the applicability is concerned, the SMGBM models are superior to the Markovian counterpart for its greater flexibility in fitting the inter transition times. Since such flexibility in model fitting directly leads to the improvements in derivative pricing, the statistical comparison between the MMGBM and SMGBM samples becomes particularly important. For the comparison purpose, we propose a discriminating statistics whose sampling distribution varies drastically, under the regime switching assumption, with varying choices of instantaneous rate parameter. Our discriminating statistics is constructed using some descriptive statistics of squeeze duration of Bollinger band, which seems to be the most natural approach. The sampling distribution of the descriptive statistics of occupation measure of Bollinger band under a particular model hypothesis need not have a nice form and thus one may not be able to identify that as one of a few known distributions. Hence the inference cannot be done analytically.

In spite of the lack of mathematical tractability, one may surely obtain empirical distribution of the statistics using a reliable simulation procedure. This is a standard approach in such circumstances. We refer \cite{JD} for further details. In Theiler et al \cite{JT} this is termed as the typical realization surrogate data approach.  The surrogate data approach is  generally perceived one of the most powerful method for testing of hypothesis. The available algorithms to generate surrogate can be categorized into two classes namely typical realizations and constrained realization (see \cite{SS1}). This paper considers the typical realization approach.

In this article, we study the geometric Brownian motion, or GBM in short, and its generalizations by using binary regime switching. By a binary regime, we mean that there exists an unobserved two-state stochastic process whose movement causes in changing the market parameters i.e. drift and volatility coefficients.  Let $S_t$ denote the asset price at time $t$, under the binary regime switching GBM. Then $\{S_t\}_{t\geq 0}$ satisfies $S_0>0$ and
\begin{align}\label{binary}
dS_t=S_t\left(\mu(X_t)\,dt+\sig(X_t)\,dB_t\right),
\end{align}
where  $\{X_t\}_{t\geq 0}$  is a $\{1,2\}$-valued stochastic process and $\mu(X_t),\sig(X_t)$ are the drift and the volatility coefficients and $B_t$ is the standard Brownian motion. In general $\{X_t\}_{t\geq 0}$ is chosen to be Markov or semi-Markov process. The GBM model is a special case of this where $\mu(1)=\mu(2)$ and $\sig(1)=\sig(2)$. Since the GBM model is quite popular among derivative market practitioners, we have compared the distribution of the discriminating statistics for a surrogate data produced from geometric Brownian motion (GBM) with that for an empirical data. We also have compared the sampling distributions of statistics for the surrogate data produced from Markov modulated GBM and semi-Markov modulated GBM respectively. We have considered eighteen sectoral indices from Indian stock market for this purpose. Corresponding to each time series data, we have first figured out the sampling distributions of the test statistics under various null hypotheses. In this paper, a systematic comparison of the proposed test statistics is carried out for each for each of the GBM, MMGBM, SMGBM cases. As an outcome, the GBM hypothesis is rejected for each time series data with nearly 100\% confidence. The MMGBM hypotheses are also rejected with more than 95\% confidence for some time series. However, the SMGBM hypothesis could not be rejected at such confidence level.

The rest of this article is arranged in the following manner. We dedicate the second section in describing the statistics based on the squeeze of the Bollinger bands. In this section we also recall some definitions from statistics and introduce the notion of squeeze durations in our context. The empirical study has been done in Section 3. The sections 4 and 5 contain concluding remark and acknowledgments respectively.
\section{Discriminating statistics based on Squeeze Duration }
In this section, a discriminating statistics is proposed whose sampling distribution varies drastically, under the regime switching assumption, with varying values of instantaneous rate parameter. The discriminating statistics is taken as vector valued where every component is a descriptive statistics of squeeze duration of Bollinger band. This section is dedicated in describing the statistics and the numerical methods for obtaining its sampling distributions. The actual numerical experiments are deferred to the next section. This section is organized in four subsections.
\subsection{Bollinger Band}\label{subsec:bollband} Keltner channel and Bollinger bands based on the empirical volatility are the most popular indicators for trading.
John Bollinger introduced the concept of Bollinger band for pattern recognition in 1980s. Bollinger bands provide a time varying interval for any financial time series data. The end points of the intervals are computed based on the moving average and the moving sample standard deviation of the past data of fixed window size. Now we present a formal definition of the Bollinger bands of an asset.
\begin{Definition}\label{BB}
	A Bollinger band of a given time series data consists of three lines on the time series plot, computed based on immediate lag values of fixed length $n$ say. The middle line is the moving average of the time series with window size $n$. The upper and the lower lines are exactly $k\sig$ unit away from the middle line where $k$ is a fixed constant and $\sig$ is the sample standard deviation obtained from the last $n$ numbers of lag values.
\end{Definition}
It is important to note that the main focus of the Bollinger band is to capture the fluctuation, to be more precise, the volatility coefficient of the time series. Hence the closeness of the upper and lower line is termed as squeeze, and is the indication of low volatility of a particular time series. On the contrary, when the boundaries of band are far from each other, that corresponds to a high volatility. For more details about Bollinger bands and squeeze we refer to \cite{bollinger}.
\subsection{$p$-Squeeze Durations}
In this paper, we consider the Bollinger bands of the simple return of a financial time series. We introduce some important notations and definitions which would be used subsequently. Let $S=\{S_k\}_{k=1}^N$ denote an equispaced financial time series. The simple return of $S$ is defined by
\begin{align}\label{simple-ret}
r_{k} := \dfrac{S_{k}-S_{k-1}}{S_{k-1}},~~k=2,3,\ldots,N.
\end{align}
\begin{Definition}[$\hat{\mu},\hat{\sig}$]\label{mu:sighat}
	By fixing the window size as $n$, the moving average $\{m_k\}_{k=n+1}^N$ and the sample standard deviation $\{\sig_k\}_{k=n+1}^N$ are given by
	\begin{align}
	m_k &:= \frac{1}{n}\sum_{i=0}^{n-1} r_{k-i},\label{movavg}\\
	\sig_k &:= \sqrt{\dfrac{1}{n-1} \sum_{i=0}^{n-1} \left(r_{k-i}\right)^2 - \dfrac{n}{n-1} m_k^2}\label{movstd},
	\end{align}
	for $k\geq n+1$. The empirical volatility $\hat{\sig} = \{\hat{\sig}_k\}_{k= n+1}^N$ is given by $\hat{\sig}_k := \dfrac{\sig_k}{\sqrt{\Delta}}$ for all $k \geq n+1 $, where $\Delta$ is the length of the time step in year unit. Similarly, the empirical drift $\hat{\mu}=\{\hat{\mu}_k\}_{k= n+1}^N$ is given by $\hat{\mu}_k := \dfrac{m_k}{\Delta}$ for all $k\geq n+1$.
\end{Definition}

\begin{Definition}\label{ecdf}
	Let $y=\{y_k\}_{k=1}^m$ be a random sample of a real valued random variable. Then the empirical cumulative distribution function or ecdf $\hat{F}_y$ is defined as
	\begin{equation*}
	\hat{F}_y(x):=\frac{1}{m}\ds\sum_{k=1}^{m}\mathds{1}_{[0,\infty)}(x-y_k),
	\end{equation*}	
	where given a subset $A$, $\mathds{1}_{A}$ denotes the indicator function of $A$.
\end{Definition}
\begin{Definition}[$p$-percentile]
	Let $\hat{F}_y$ be the ecdf of $y=\{y_k\}_{k=1}^m$. Then for any $p\in (0,1)$, the $p$-percentile of $y$, denoted by $\hat{F}_y^{\leftarrow}(p)$, is defined as
	\begin{equation*}
	\hat{F}_y^{\leftarrow}(p):=\inf\bigg\{x\big|\hat{F}_y(x)\geq p\bigg\}.
	\end{equation*}
\end{Definition}
For mathematical tractability we use a particular percentile of $\hat{\sig}$ as threshold in defining the squeeze of the Bollinger band.
\begin{Definition}[$p$-squeeze] Given a $p\in (0,1)$, an asset is said to be in $p$-squeeze at $k$-th time step if the empirical volatility $\hat{\sig}_k$, as defined above, is not more than $\hat{F}_{\hat{\sig}}^{\leftarrow}(p)$.
\end{Definition}
We introduce the sojourn times of the $p$-squeeze below.
\begin{Definition}\label{SqD}
	For a fixed $p \in (0,1)$ and a given time series $\{S_k\}_{k=1}^N$, let $\{(a_i,b_i)\}_{i=1}^\infty$ be an extended real valued double sequence given by \begin{equation*}
	\left\{ \begin{array}{lll}
	a_0=0\\
	b_{i-1}:=\min\{k\geq a_{i-1}|\hat{\sig}_{k}>\hat{F}_{\hat{\sig}}^{\leftarrow}(p)\}\\
	a_i:=\min\{k\geq b_{i-1}|\hat{\sig}_{k}\leq \hat{F}_{\hat{\sig}}^{\leftarrow}(p)\},
	\end{array}\right.
	\end{equation*}
	for $i=1,2,\ldots$ and by following the convention of $\min\emptyset =+\infty$, where $\hat{\sig}$ is as in Definition \ref{mu:sighat}. Then the sojourn time durations for the p-squeezes are $\{d_i\}_{i=1}^L$, where  $d_i:=b_i-a_i$ and $L:=\max\{i|b_i<\infty\}$, provided $L\geq 1$.
\end{Definition}

We note that one must multiply each $d_i$ by $\Delta$ to obtain the squeeze durations in year unit. We would consider the finite sequence $\{d_i\}_{i=1}^L$ as a single object. In particular, we call $d_i$ as the $i$-th entry of the $p$-squeeze duration or $p$-SqD in short for the given time series $\{S_k\}_{k=1}^N$. We call $L$ to be the length of $p$-SqD.
\begin{Remark}\label{errorecdf}
	In a reasonably large and practically relevant time series data, the length of $p$-SqD is considerably small. Hence a non parametric estimation of the entries of $p$-SqD using empirical cdf is not practicable as that would have a high standard error. Hence, only a collection of some descriptive statistics such as mean($\bar{d}$), standard deviation($s$), skewness($\nu$), kurtosis($\kappa$) of $p$-SqD can reliably be obtained and compared.
\end{Remark}
\subsection{A Discriminating Statistics}
We first consider a discrete time version of continuous time theoretical asset price model with the time step identical to that of the time series data. We note that, for that theoretical model, the corresponding $p$-SqD is a random sequence with random length. However, the corresponding descriptive statistics as above would constitute a random vector of fixed length whose sampling distribution would be sought for. A comparison of $(\bar{d}, s, \nu, \kappa)$ with respect to that sampling distribution would be the central idea for statistical inference. However, this should not lead to the only criterion for rejecting a model. Of course there are many other natural criteria for the same. Those criteria are typically considered as constraints on parameterization of the class of models of our interest. Here we illustrate with an example. If we restrict ourselves to all possible MMGBM models with two regime as in \cite{BU}, then
\begin{align}\label{thetacls}
\Theta=\{(\mu(1),\sig(1),\la_1,\mu(2),\sig(2),\la_2)|\mu(i)\in\mathbb{R},\sig(i)>0,\la(i)>0, i=1,2\}
\end{align}
is the class of parameters, where $\mu(i)$ and $\sig(i)$ are the drift and the volatility coefficients respectively and $\begin{pmatrix}
-\la_1 & \la_1\\
\la_2 & -\la_2
\end{pmatrix}$ is the rate matrix for the Markov chain. Therefore the estimation problem boils down to a constrained minimization on a set $\mathscr{C}\subset\Theta$ of the following functional $f:\Theta\to\mathbb{R}$ given by
\begin{align}\label{obj:func}
f(\theta)=E^\theta\left[(\bar{d}^\theta-\bar{d})^2+(s^\theta-s)^2+(\nu^\theta-\nu)^2+(\kappa^\theta-\kappa)^2\right],
\end{align}
where $(\bar{d}^\theta,s^\theta,\nu^\theta,\kappa^\theta)$ is the descriptive statistics vector of a member with parameter $\theta\in \Theta$, provided the minimizer exists.
%In view of Remark \ref{errorecdf}, instead of non-parametric estimator, we propose a parametric estimator. Let $\bar{d}$, $s$, $\nu$, $\kappa$ denote the four descriptive statistics mean, standard deviation, skewness and kurtosis respectively of $p$-SqD of the original data. In order to fit the distributional properties using simulation, we look for a parametric class of binary regime switching models $\Theta$. Since our investigation is on binary regime assumption,
%the unknown parameters are the drift, the volatility and the transition rate for each regime. Therefore $\Theta$ can be viewed as

The main difficulty in taking $\mathscr{C}=\Theta$ is its time complexity due to a large scope of parameter values. We introduce a fixed set of constraints, $p$-admissible class($\mathscr{C}_p$-class), which is a subclass of all possible regime switching models.
\begin{Definition}[$\mathscr{C}_p$-class]\label{padcls}
	Given a time series data and a fixed $p\in (0,1)$, a regime switching model is said to be in $\mathscr{C}_p$-class of models if the model satisfies the following properties.
	\begin{itemize}
		\item[i.] The long run average of drift coefficient matches with the time average of empirical drift $\hat{\mu}$.
		\item[ii.] The long run average of volatility parameters matches with the time average of empirical volatility $\hat\sig$.
		\item[iii.] The long run proportion of time that the volatility process stays below $\hat{F}_{\hat{\sig}}^{\leftarrow}(p)$ is $p$, provided the volatility process is not constant.
	\end{itemize}	
\end{Definition}

In view of Remark \ref{errorecdf}, we construct a discriminating statistics $\mathbf{T}=(T_1,T_2,\ldots,T_r)$ using $r$ number of descriptive statistics of the $p$-SqD. To be more specific we choose $T_1:=\frac{1}{L}\ds\sum_{i=1}^{L}d_i$,
$T_2:=\sqrt{\frac{1}{L-1}\ds\sum_{i=1}^{L}(d_i-T_1)^2}$, $T_3:=\dfrac{\frac{1}{L}\ds\sum_{i=1}^{L}(d_i-T_1)^3}{T^3_2}$, $T_4:=\dfrac{\frac{1}{L}\ds\sum_{i=1}^{L}(d_i-T_1)^4}{T^4_2}$ etc. Although our test statistics is based on squeeze durations which are amenable to capture the sojourn times of regime transitions but it is not at all obvious that it would indeed be successful to capture those unobserved switchings. The main difficulty lies in the fact that a larger moving window size ($n$) in defining $\hat{\sig}$ ignores more number of intermittent transitions and a smaller window size corresponds to higher standard error. So far window size is concerned there is a popular choice of window size by practitioners, i.e., $n=20$ for computing the empirical volatility. In view of these, we fix $n=20$ now onward in the definition of $\mathbf{T}$.

Next we describe the procedure, adopted in this paper, of obtaining the sampling distribution of $\mathbf{T}$ under binary regime switching model hypothesis.
\subsection{Sampling distribution of the statistics}\label{tetspres}
% define binary regime switching model
In this section we give a detailed description of numerical computation of sampling distribution of $\mathbf{T}$ statistics under the null hypothesis using Monte Carlo method, which is popularly known as typical surrogate approach following \cite{JT}. It is important to note that the hypothesis testing, relevant to us, is of composite type (see \cite{JD}). The main purpose is to test a meaningful composite null hypothesis. The procedure is as follows
\begin{itemize}
	\item[(a)] Given a time series $S$, the $p$-admissible class $\mathscr{C}_p$ under the null hypothesis is identified. A non-empty subclass $\mathscr{A}$ of $\mathscr{C}_p$ is fixed.
	\item[(b)] For each $\theta \in \mathscr{A}$, $B$ number of time series $\{X^1,X^2,\ldots,X^B\}$ are simulated from the corresponding model $\theta$ with the same time step as in $S$. We call these, the surrogate data of $S$ corresponding to $\theta$.
	\item[(c)]  Let $\mathbf{t}^*:=\mathbf{T}(S)$ be the value of $\mathbf{T}$ of the observed data $S$ and $\mathbf{t}:=\{\mathbf{t}^1,\mathbf{t}^2,\ldots,\mathbf{t}^B\}$ be the values of $\mathbf{T}$ for the surrogate data $\{X^1,X^2,\ldots,X^B\}$, where $\mathbf{t}^i=(t^i_1,t^i_2,\ldots,t^i_r)=\mathbf{T}(X^i)$ .
	\item[(d)]  By keeping a two sided test in mind we define $\alpha^\theta_r$ in the following manner
   $$\alpha^\theta_r:=2\ds\min_{j\leq r}\,g_B\left(\ds\sum_{i=1}^{B}\mathds{1}_{[0,\infty)}(t^*_j-t_j^i)\right),$$ where $g_B(x):=\frac{x\wedge(B-x)}{B}$, and $\mathbf{t}^*=(t^*_1,t^*_2,\ldots,t^*_r)$.
	\item[(e)] Therefore the $\alpha$-value for the test of the class $\mathscr{A}$ is given by
	\[\alpha_r=\ds\max_{\theta\in \mathscr{A}}\,\alpha_r^\theta.\]
	\item[(f)] We reject the hypothesis that $S$ is a sample from a model in the class with confidence $100(1-\alpha_r)\%$, provided $\alpha_r$ is reasonably small.
\end{itemize}

\begin{figure}[h]
	\begin{center}
		\includegraphics[width=0.5\textwidth]{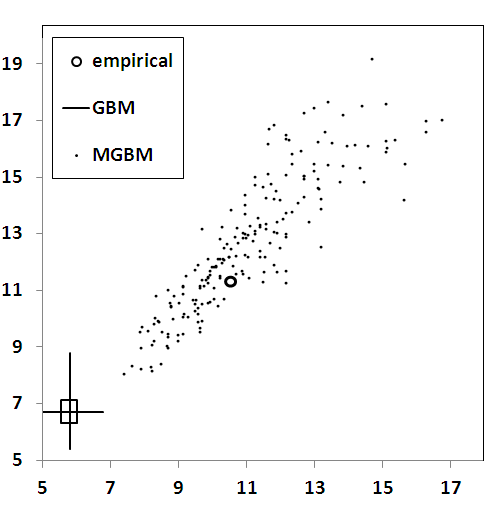}
		
	\end{center}
	
	\caption{Plot of $(T_1,T_2)$ for observation and various different surrogate data.}
	\label{fig1}
\end{figure}

\begin{Remark}
	It is important to note, the above method has a pathetic limitation due to the ``curse of dimensionality." Or in other words for a given model $\theta$, the probability of observing the value of $\alpha^\theta_r$ to be smaller than a very small value is not so small when $r$ is large. But it is well known that the curse is not so fatal for the dimension $r$ less than five. Therefore we restrict ourselves in four dimensional testing.
\end{Remark}
For the purpose of illustration here we consider a specific time series $S$ and $r=2$ i.e., the test statistics $\mathbf{T}=(T_1,T_2)$. In the following plot Figure \ref{fig1}, the values $T_1$ and $T_2$ are plotted against the horizontal and vertical axes respectively. The position of the point $t^*=\mathbf{T}(S)$ is denoted as a circle in the plot. Under the null hypothesis of GBM, the $\mathscr{C}_p$-class turns out to be singleton (see Section \ref{sec:GBM} for details). The sampling distribution of $\mathbf{T}$ under the null hypothesis is computed by setting $B=200$ and that is presented using a two dimensional box plot. Furthermore, under the null hypothesis of MMGBM, followed by fixing $B=1$, and a subclass
$\mathscr{A}:=\{\frac{1}{\la_1}\in (\frac{1}{20}\mathbb{N})\cap [5,15], \sig(1)=\hat{F}_{\hat{\sig}}^{\leftarrow}(p)\}$ of $\mathscr{C}_p$, the values of $\mathbf{T}$ are plotted as dots.

In the following section, we implement the ideas developed here for more specific choice of models and discuss the implementation issues in details.

\section{Empirical study}\label{empirical}
For empirical study, we consider, $5$-minute data of several Indian stock indices from $1$-st December, 2016 to $30$-th June, 2017. Assuming there are $250$ trading days in a year and $6$ hours of trading in each day, we set  $\Delta = \frac{5}{250\times 360} \approx 5.5\times10^{-5}$. We fix $p=15\%$ throughout this section. The components of $t^*$ for each index data are given in the Table \ref{Table1} below. Every row of the table corresponds to an index, which is mentioned in the second column with their id's in the first column. The third column gives the value of $L$, the number of observations of $p$-squeeze duration of each index data.

\begin{table}[h]
	\centering
	\caption{Values of discriminating statistics $(p=15\%)$ of a $5$ -minutes data of Indian indices during 1st Dec 2016 to 30th June 2017} \label{Table1}
	\small
	
	\begin{tabular}{llrrrrr}
		
		Data &	Indices &   Occurence &    $t^*_1$ &     $t^*_2$ &  $t^*_3$ & $t^*_4$\\
		\hline
		I01 &  NIFTY 100          &        159 &  10.52 &  11.31 &    1.17 &    3.41 \\
		I02 &  NIFTY 200          &        160 &  10.45 &  11.18 &    1.29 &    3.79 \\
		I03 &  NIFTY 50           &        155 &  10.78 &  11.00 &    1.08 &    3.28 \\
		I04 &  NIFTY 500         &        152 &  11.01 &  11.40 &    1.20 &    3.63 \\
		I05 &  NIFTY BANK         &        159 &  10.52 &  11.62 &    1.39 &    4.03 \\
		I06 &  NIFTY COMMODITY   &        169 &   9.89 &  10.49 &    1.47 &    4.59 \\
		I07 &  NIFTY ENERGY       &        168 &   9.96 &  11.44 &    1.59 &    4.80 \\
		I08 &  NIFTY FIN. SER.    &        168 &   9.95 &  10.72 &    1.46 &    4.39 \\
		I09 &  NIFTY FMCG         &        178 &   9.40 &  10.15 &    1.58 &    5.01 \\
		I10 &  NIFTY INFRA        &        174 &   9.61 &  11.70 &    1.72 &    5.41 \\
		I11 &  NIFTY IT           &        159 &  10.52 &  11.36 &    1.19 &    3.35 \\
		I12 &  NIFTY MEDIA        &        173 &   9.66 &   9.49&    1.19 &    3.77 \\
		I13 &  NIFTY METAL        &        188 &   8.89 &  10.53 &    1.92 &    6.52 \\
		I14 &  NIFTY MNC          &        178 &   9.40 &  10.63 &    1.54 &    4.67 \\
		I15 &  NIFTY PHARMA       &        175 &   9.56 &  11.13 &    1.59 &    4.69 \\
		I16 &  NIFTY PSE          &        148 &  11.29 &  12.73 &    1.28 &    3.76 \\
		I17 &  NIFTY REALTY       &        177 &   9.45 &  10.49 &    1.83 &    6.02 \\
		I18 &  NIFTY SERVICE SEC. &        172 &   9.72 &  11.13 &    1.33 &    3.77\\
		\hline
	\end{tabular}
\end{table}

\subsection{GBM hypothesis}\label{sec:GBM}
Let $(\Omega,\mathcal{F},\mathbb{P})$ be a probability space on which $\{W_t\}_{t\geq 0}$ be a Brownian motion. The stock price, modeled by a geometric Brownian motion (or GBM in short) is given by
\begin{align}\label{bsm}
dS_t=S_t\left(\mu\,dt+\sig\,dW_t\right)~t\geq 0, ~ S_0>0,
\end{align}
where $\mu$ and $\sig$ are the drift and the volatility coefficients respectively. Equation \eqref{bsm} has a strong solution of the form
\begin{align}\label{solbsm}
S_t=S_0\exp\left(\mu t-\frac{1}{2}\sig^2 t+\sig W_t\right),~t\geq 0.
\end{align}
\subsubsection{Surrogate Data}
It is important to note that, the $\mathscr{C}_p$-class is singleton as $\mu$ and $\sig$ are, by using Definition \ref{padcls} (i)-(ii), $\mu=\bar{\hat{\mu}}$ and $\sig=\bar{\hat{\sig}}$, where the bar sign represents the time average.

Let $S_0$ be the initial price of a stock consisting of $N$ number of data points. Let $\{0=t_1<t_2<\ldots<t_N\}$ be a partition of time interval of the observed data series, where $t_{i+1}-t_i=\Delta$ for $i=1,2,\ldots,N-1$ and $\Delta$ be the length of time step in year unit. Then the $\mathscr{C}_p$-class of surrogate GBM can be generated by using the discretized version of \eqref{solbsm} which is given by
\begin{align}\label{dissolbsm}
S_{t_{i+1}}=S_{t_i}\exp\left((\bar{\hat{\mu}} -\frac{1}{2}\bar{\hat{\sig}}^2)\Delta +\bar{\hat{\sig}}\,Z_i\right),~ S_{t_1}=S_0
\end{align}
where $\{Z_i\mid i=1, \ldots, N-1\}$ are independent and identically distributed (i.i.d.) normal random variables with mean $0$ and variance $\Delta$. We further emphasize that the step size $\delta$, used for the discretization is identical to the step size in the observed time series. We use this convention throughout this paper.
\subsubsection{Testing of hypothesis}
We intend to test whether the value of $\mathbf{T}$ of observed index prices are outliers of $\mathbf{T}$ values coming from GBM models. For each index in Table \ref{Table1}, we set our null hypothesis,
\begin{align*}
H_0:\text{the time series is in $\mathscr{C}_p$-class of GBM.}
\end{align*}
%\footnote{Codes are mostly developed using Python and C programming languages (Cython). It will be available on request to the corresponding author. }
We again recall that the $\mathscr{C}_p$-class is indeed singleton. The following figures illustrate results from all 18 indices. Figure \ref{gbm:mean} plots $T_1$ and Figure \ref{gbm:sd} plots $T_2$ only. Each box plot is obtained by simulating the GBM model from $\mathscr{C}_p$-class $200$ times. The triangle plots are the representative for original data of all the indices. Here we see that the triangles appear far from the box plots.

\begin{minipage}{\linewidth}
	\centering
	\begin{minipage}{0.48\linewidth}
		\begin{figure}[H]
			\includegraphics[width=\linewidth]{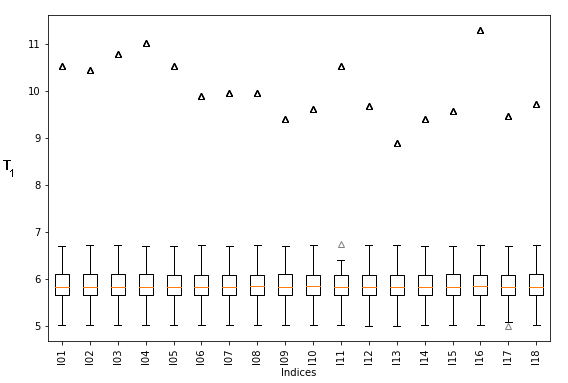}
			\caption{ Sampling distribution of $T_1$ under GBM hypothesis}\label{gbm:mean}
		\end{figure}
	\end{minipage}
	\hspace{0.02\linewidth}
	\begin{minipage}{0.48\linewidth}
		\begin{figure}[H]
			\includegraphics[width=\linewidth]{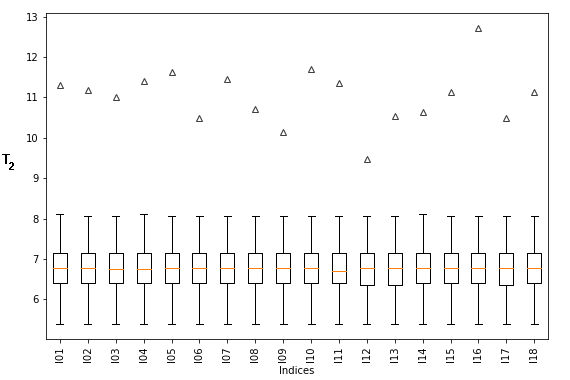}
			\caption{ Sampling distribution of $T_2$ under GBM hypothesis}\label{gbm:sd}
		\end{figure}
	\end{minipage}
\end{minipage}

Therefore the above Figures \ref{gbm:mean} and \ref{gbm:sd} indicate a strong rejection for the null hypothesis of GBM model. We continue our investigation with binary regime switching Markov modulated geometric Brownian motion in the following subsection.
\subsection{Markov modulated GBM hypothesis}
In this subsection, we study the testing of hypothesis of $\mathscr{C}_p$-class of MMGBM assumption with binary regimes. In this case the equation \eqref{binary} is dependent on a two state Markov process $\{X_t\}_{t\geq 0}$. It is important to note that a continuous time Markov chain can be characterized by its instantaneous transition rate matrix.

\subsubsection{Surrogate Data}
In this subsection, we restrict our investigation in a particular subclass $\mathscr{A}$ of $\mathscr{C}_p$ which are embedded in
\begin{align}\label{class:mmgbm}
\Theta=\{\theta=(\mu(1),\sig(1),\la_1,\mu(2),\sig(2),\la_2)|\mu(i)\in\mathbb{R},\sig(i)>0,\la_i>0, i=1,2\}.
\end{align}

Here the transition rate matrix for the Markov chain is given by \[
\Lambda:=
\begin{pmatrix}
-\la_1 & \la_1\\
\la_2 & -\la_2
\end{pmatrix}.
\]
Therefore the sojourn time distribution of state $i$ is Exp$(\la_i)$ for $i=1,2$. Now since $\mathscr{A}\subset \mathscr{C}_p\subset \Theta$, by using Definition \ref{padcls}(iii), we have
\begin{equation}\label{lap1}
\dfrac{\frac{1}{\la_1}}{\frac{1}{\la_1}+\frac{1}{\la_2}}=p.
\end{equation}
Using Definition \ref{padcls}(i), \eqref{lap1} the drift coefficients $\mu(i)$ satisfy the following relation
\begin{equation}\label{rel:mu}
p\,\mu(1)+(1-p)\,\mu(2)=\bar{\hat{\mu}}.
\end{equation}
Also using Definition \ref{padcls}(ii) and \eqref{lap1} the volatility coefficients $\sig(i)$ have the relation below
\begin{equation}\label{rel:si}
p\,\sig(1)+(1-p)\,\sig(2)=\bar{\hat{\sig}}.
\end{equation}
After a simple computation \eqref{lap1} becomes
\begin{equation}\label{lap}
\la_1=\left(\frac{1}{p}-1\right)\la_2.
\end{equation}
Thus $\mathscr{C}_p\subset \Theta$ is the set of six parameters satisfying equations \eqref{rel:mu}, \eqref{rel:si} and \eqref{lap}. We choose $\mathscr{A}$ by fixing $\mu(1)=\mu(2)$ and $\sigma(1)=\hat{F}_{\hat{\sig}}^{\leftarrow}(p)$. Thus $\mathscr{A}$ is a subset of the solution space of five equations in six unknowns, or in other words, $\mathscr{A}$ can be viewed as a one-parameter family of models. Next to generate surrogate data, we need to discretize the MMGBM model corresponding to each member of $\mathscr{A}$. To this end we discretize \eqref{binary} and perform Monte Carlo simulation.
The discretization of MMGBM surrogate of the form \eqref{binary} is given by
\begin{align}\label{dissolmbsm}
\no S_{t_{i+1}} & =  S_{t_i}\exp\left(\left(\mu(X_i) -\frac{1}{2}\sig^2(X_i)\right)\Delta +\sig(X_i)\,Z_i\right),\\
X_{i+1} & = X_i-(-1)^{X_i}\,P_i,
\end{align}
where $\{P_i \mid i=1, \ldots, N-1\}$ are independent to $Z_j$ for all $j$ and for each given $X_i$, the conditional distribution of $P_i$ is independent of $\{P_1, P_2, \ldots, P_{i-1}\}$ and follows $Bernoulli(\lambda_{X_i}\Delta )$, a Bernoulli random variable with $Prob( P_i=1\mid X_i)=\lambda_{X_i}\Delta $, provided $\Delta \ll \min \{1/\lambda_{i} \mid i=1,2\}$. Here for each $i$, $Z_i$ is as in \eqref{dissolbsm}. For sufficiently small $\Delta$, the above discrete model approximates the continuous time MMGBM model. A proof of this fact is excluded here.

\subsubsection{Testing of hypothesis}
We intend to test whether the values of $\mathbf{T}$ of observed index prices are outliers of $\mathbf{T}$ values coming from MMGBM models. For each index in Table \ref{Table1}, we set our null hypothesis,
\begin{align*}
H_0:\text{the time series is in the class $\mathscr{A}$ of MMGBM.}
\end{align*}
To test $H_0$, we adopt the typical realization surrogate data approach and consider the discrete model \eqref{dissolmbsm}. For every $\theta\in \mathscr{A}$, we perform Monte Carlo simulation two hundred times $(B=200)$. Then we record the $\alpha_1$, $\alpha_2$, $\alpha_3$ and $\alpha_4$ values for each index in Table \ref{Table2}.
In order to enhance understanding of the composite hypothesis, here we consider the first time series I01 to illustrate the sampling distribution of $\mathbf{T}$ for some models in $\mathscr{A}$ using the following four 2-D projection plots. In each of the Figures \ref{mgbm:t1t2}-\ref{mgbm:t1t4}, the circle plot represents $t^*$, the $\mathbf{T}$ value of I01. Furthermore, there are three two-dimensional box plots corresponding to $\frac{1}{\la_1}$ equal to $5,10$ and $15$ respectively. By abuse of notation we write $\frac{1}{\la_1}$ as $\theta$. It seems from Figures \ref{mgbm:t1t2} that, the least square estimate of $\frac{1}{\la_1}$ should lie in between $5$ and $10$, so far $(T_1,T_2)$ is concerned. Indeed as in Table \ref{Table2}, the $\alpha_2$ value (0.395), which depends only on $T_1$ and $T_2$, is considerably large. However, Figure \ref{mgbm:t3t4} implies that the least square estimate of $\frac{1}{\la_1}> 15$ by considering $(T_3,T_4)$ only. On the other hand, each of Figures \ref{mgbm:t1t3} and \ref{mgbm:t1t4} implies that the circle plot is an outlier. Thus no combination of parameters in the composite hypothesis can explain the observed data. We can also observe another very important feature from these four plots. The sampling distributions of the statistics change monotonically with the change of parameter $\theta$. Such monotonic sensitivity of the statistics is crucial for successful inference.

From Table \ref{Table2} it is evident that $H_0$ can be rejected for each index with $\alpha$ value $5\%$ or smaller. Finally, since all the $\alpha_4$ are less than $5\%$, in the MMGBM column of Table \ref{Table2}, we reject the null hypothesis with $95\%$ confidence for each index data.

\begin{minipage}{\linewidth}
	\centering
	\begin{minipage}{0.48\linewidth}
		\begin{figure}[H]
			\includegraphics[width=\linewidth]{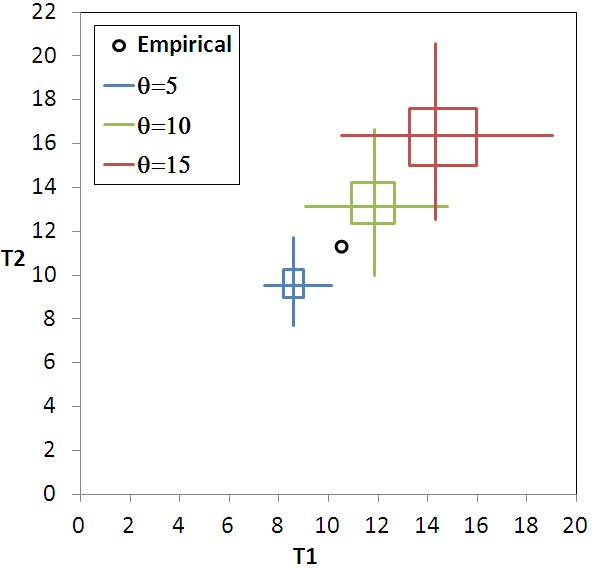}
			\caption{ $T_1$ and $T_2$ under MMGBM hypothesis}\label{mgbm:t1t2}
		\end{figure}
	\end{minipage}
	\hspace{0.02\linewidth}
		\begin{minipage}{0.48\linewidth}
		\begin{figure}[H]
			\includegraphics[width=\linewidth]{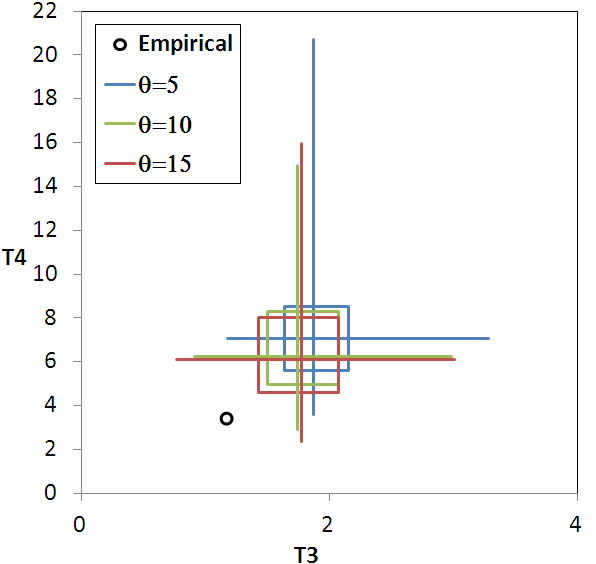}
			\caption{ $T_3$ and $T_4$ under MMGBM hypothesis}\label{mgbm:t3t4}
		\end{figure}
	\end{minipage}
\end{minipage}

\begin{minipage}{\linewidth}
	\centering
	\begin{minipage}{0.48\linewidth}
		\begin{figure}[H]
			\includegraphics[width=\linewidth]{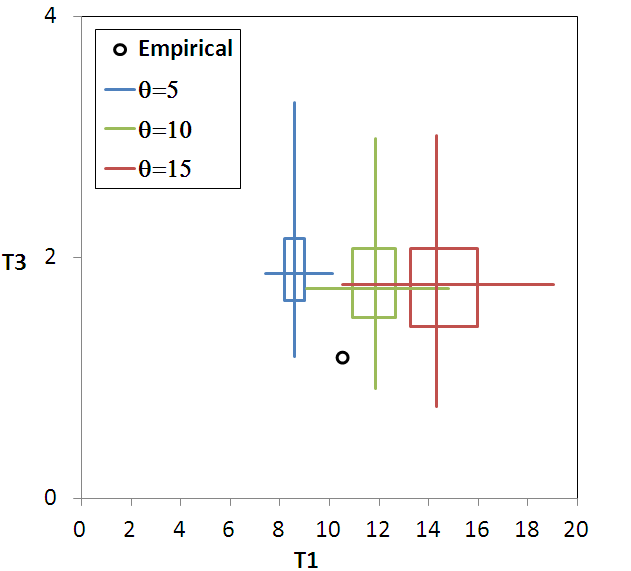}
			\caption{ $T_1$ and $T_3$ under MMGBM hypothesis}\label{mgbm:t1t3}
		\end{figure}
	\end{minipage}
    \hspace{0.02\linewidth}
	\begin{minipage}{0.48\linewidth}
		\begin{figure}[H]
			\includegraphics[width=\linewidth]{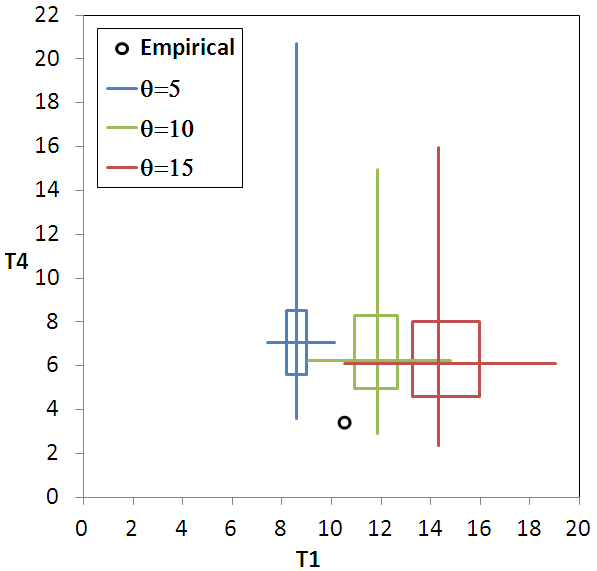}
			\caption{ $T_1$ and $T_4$ under MMGBM hypothesis}\label{mgbm:t1t4}
		\end{figure}
	\end{minipage}
	
\end{minipage}

\subsection{Semi-Markov modulated GBM} In this subsection, we consider the SMGBM assumption with binary regimes. In this case the equation \eqref{binary} is dependent on a two state semi-Markov process $\{X_t\}_{t\geq 0}$. A semi-Markov process can be characterized by its instantaneous transition rate function. Alternatively a semi-Markov process can also be specified by fixing the transition probability matrix of the embedded chain and the conditional distribution of holding time at each state. 
\subsubsection{Surrogate Data}
A subclass of the following all class of models
\begin{align*}
\Theta=\{\theta=(\mu(1),\sig(1),\la_1(\cdot),\mu(2),\sig(2),\la_2(\cdot)|\mu(i)\in\mathbb{R},\sig(i)>0,~\la_i(\cdot)>0,~ i=1,2\}.
\end{align*}
would be considered now. For every $\theta$ in $\mathscr{C}_p\subset\Theta$, $\mu(i)$ and $\sig(i)$ satisfy the equation \eqref{rel:mu} and \eqref{rel:si}. The transition rate matrix for the semi-Markov chain is a matrix valued function on $[0,\infty)$, given by \[
\Lambda(y):=
\begin{pmatrix}
-\la_1(y) & \la_1(y)\\
\la_2(y) & -\la_2(y)
\end{pmatrix}~~\forall~ y\in [0,\infty).
\]
Now for illustration purpose, $\mathscr{A}$ is chosen in the following manner. The holding time distribution of the state $i$ is $\Gamma(k_i,\la_i)$ for $i=1,2$, where $\Gamma(k_i,\la_i)$ denote the gamma distribution with shape $k_i$ and rate $\la_i$. Then it follows from \cite{GhS} that $\la_i(y)$ is the hazard rate of $\Gamma(k_i,\la_i)$ and is given by
$\la_i(y) = \frac{\la_i^{k_i}y^{k_i-1}e^{-\la_iy}}{\Gamma(k_i)-\gamma(k_i,\la_iy)}$, where $\gamma$ is the lower incomplete gamma function. Since the expectation of $\Gamma(k_i,\la_i)$ is $\frac{k_i}{\la_i}$, it follows from Definition \ref{padcls}(iii), that
\[\frac{\frac{k_1}{\la_1}}{\frac{k_1}{\la_1}+\frac{k_2}{\la_2}}=p,\]
i.e. \[\frac{k_2}{\la_2}=\left(\frac{1}{p}-1\right)\frac{k_1}{\la_1}.\]
In addition to these, as before, we further assume that $\mu(1)=\mu(2)$, $\sigma(1)=\hat{F}_{\hat{\sig}}^{\leftarrow}(p)$ and $k_1=k_2$. Thus $\mathscr{A}$ is the solution space of six equations in eight unknowns or in other words $\mathscr{A}$ is a two parameter subfamily of $\Theta$. For drawing samples from each member of $\mathscr{A}$ using Monte Carlo simulation, we first discretize \eqref{binary}.
The discretization scheme for SMGBM surrogate of \eqref{binary} is given by
\begin{align}\label{dissolsbsm}
\no S_{t_{i+1}} & =  S_{t_i}\exp\left(\left(\mu(X_i) -\frac{1}{2}\sig^2(X_i)\right)\Delta +\sig(X_i)\,Z_i\right),\\
\no X_{i+1} & = X_i+(-1)^{X_i}\,P_i,\\
Y_{i+1} & = \left(Y_i+i\Delta\right)\left(1-P_i\right),
\end{align}
where $\{Z_i\}_i$ are as in \eqref{dissolbsm} and $\{P_i \mid i=1, \ldots, N-1\}$ are independent to $Z_j$ for all $j$ and for each given pair $(X_i, Y_i)$, the conditional distribution of $P_i$ is independent of $\{P_1, P_2, \ldots, P_{i-1}\}$ and follows $Bernoulli(\lambda_{X_i}(Y_i)\Delta )$, a Bernoulli random variable with $Prob( P_i=1\mid X_i, Y_i)=\lambda_{X_i}(Y_i) \Delta $, provided $\Delta \ll \min \{1/\lambda_{i}(y) \mid y\ge 0, i=1,2\}$. This discretization is obtained from the semi-martingale representation of the semi-Markov process, as in \cite{GhS}. The readers are referred to \cite{GhS} for more details about this representation of semi-Markov process.

\subsubsection{Testing of hypothesis}
We set our null hypothesis for all index
\begin{align*}
H_0:\text{the time series is in the class $\mathscr{A}$ of SMGBM.}
\end{align*}
From Table \ref{Table2} below, $H_0$ cannot be rejected for any index with a significant level of confidence. Hence we cannot reject the superset also. Or in other words, we cannot reject the composite hypothesis that the data, under study, is drawn from a SMGBM population.
\begin{table}[h]
	\centering
	\caption{The $\alpha$-vales for all the indices} \label{Table2}	
	\begin{tabular}{ll|rrrr|rrrr}
		\hline
		\multicolumn{2}{c}{}& \multicolumn{4}{|c|}{MMGBM}& \multicolumn{4}{|c}{SMGBM}\\
		
		{} & Index &   {$\alpha_1$} &  {$\alpha_2$} &   {$\alpha_3$} &   {$\alpha_4$} &   {$\alpha_1$} &  {$\alpha_2$} &   {$\alpha_3$} &   {$\alpha_4$}\\
		\hline
		1  &   I01 & 0.490 &  0.395 &  0.045 &  0.040 & 0.495 & 0.485 & 0.225 & 0.095 \\
		2  &   I02 & 0.490 &  0.400 &  0.050 &  0.045 & 0.500 & 0.455 & 0.195 & 0.105 \\
		3  &   I03 & 0.470 &  0.420 &  0.050 &  0.035 & 0.495 & 0.470 & 0.230 & 0.100 \\
		4  &   I04 & 0.435 &  0.395 &  0.055 &  0.040 & 0.500 & 0.455 & 0.180 & 0.105 \\
		5  &   I05 & 0.415 &  0.395 &  0.055 &  0.040 & 0.495 & 0.415 & 0.195 & 0.105 \\
		6  &   I06 & 0.465 &  0.390 &  0.055 &  0.040 & 0.485 & 0.415 & 0.200 & 0.115 \\
		7  &   I07 & 0.430 &  0.430 &  0.060 &  0.040 & 0.500 & 0.420 & 0.205 & 0.100 \\
		8  &   I08 & 0.475 &  0.420 &  0.050 &  0.035 & 0.495 & 0.455 & 0.235 & 0.100 \\
		9  &   I09 & 0.455 &  0.420 &  0.085 &  0.050 & 0.495 & 0.425 & 0.220 & 0.105 \\
		10 &   I10 & 0.460 &  0.390 &  0.055 &  0.040 & 0.485 & 0.425 & 0.200 & 0.115 \\
		11 &   I11 & 0.455 &  0.420 &  0.055 &  0.040 & 0.490 & 0.430 & 0.190 & 0.100 \\
		12 &   I12 & 0.480 &  0.395 &  0.050 &  0.050 & 0.500 & 0.470 & 0.210 & 0.115 \\
		13 &   I13 & 0.490 &  0.405 &  0.050 &  0.040 & 0.500 & 0.470 & 0.215 & 0.105 \\
		14 &   I14 & 0.430 &  0.395 &  0.055 &  0.040 & 0.485 & 0.415 & 0.195 & 0.100 \\
		15 &   I15 & 0.490 &  0.410 &  0.050 &  0.035 & 0.480 & 0.480 & 0.235 & 0.100 \\
		16 &   I16 & 0.435 &  0.395 &  0.055 &  0.040 & 0.495 & 0.425 & 0.200 & 0.105 \\
		17 &   I17 & 0.470 &  0.410 &  0.045 &  0.030 & 0.495 & 0.400 & 0.195 & 0.095 \\
		18 &   I18 & 0.425 &  0.395 &  0.055 &  0.040 & 0.490 & 0.430 & 0.205 & 0.100 \\
		\hline
	\end{tabular}
\end{table}
\section{Conclusion}
In this paper, we have developed a statistical technique to test the hypothesis of the binary regime switching extensions of GBM. The proposed statistical procedure involves four steps. In the first step we discretize the continuous time model. Monte-Carlo simulations of the discrete model is carried out with specific combination of parameters in the second step. In the third step we compute the sampling distribution of the statistics under the model hypothesis for various combinations of parameters. In the final step we compare the statistics value of the observed data with the sampling distributions obtained in the third step. Despite mathematical intractability, we present the procedure systematically with sufficient rigor. In view of the unavailability of any reported investigation on inference of regime switching models, we have taken the most natural and simplistic approach to the problem. We have successfully observed that even with this simple approach, the sampling distributions of the statistics exhibit a monotonic sensitivity to the model parameters. Thus the proposed statistics is suitable for the inference problem. The paper is written for a very broad class of readers by eliminating several technical jargons. This work, in principle, gives a framework and opens up opportunity of many research investigation concerning inference of regime switching dynamics arising in several different applied fields.

\section{Acknowledgments}
The authors are grateful to Prof. Uttara Naik Nimbalkar and Dr Kedar  Mukherjee for some very needful help. The authors also thank Sanket Nandan for some useful discussions.

\bibliographystyle{plain}

\end{document}